\documentclass[letterpaper,aps,prb,twocolumn,showpacs,groupedaddress,floatfix]{revtex4}

\usepackage{graphicx}
\usepackage[normal]{caption2}

\begin{document}

\title{Magnetic properties and magnetostructural phase transitions in
$\mathbf{Ni}_{2+x}\mathbf{Mn}_{1-x}\mathbf{Ga}$ shape memory
alloys}

\author{V.~V.~Khovailo}
\affiliation{National Institute of Advanced Industrial Science and
Technology, Tohoku Center, Sendai 983-8551, Japan}

\author{V.~Novosad}
\affiliation{Materials Science Division, Argonne National
Laboratory, Argonne, Illinois 60439}

\author{T.~Takagi}
\affiliation{Institute of Fluid Science, Tohoku University, Sendai
980--8577, Japan}

\author{D.~A.~Filippov}
\author{R.~Z.~Levitin}
\author{A.~N.~Vasil'ev}
\affiliation{Physics Faculty, Moscow State University, Moscow
119899, Russia}

\begin{abstract}
A systematic study of magnetic properties of
Ni$_{2+x}$Mn$_{1-x}$Ga ($0 \le x \le 0.19$) Heusler alloys
undergoing structural martensite-austenite transformations while
in ferromagnetic state has been performed. From measurements of
spontaneous magnetization, $M_s(T)$, jumps $\Delta M$ at
structural phase transitions were determined. Virtual Curie
temperatures of the martensite were estimated from the comparison
of magnetization in martensitic and austenitic phases. Both
saturation magnetic moments in ferromagnetic state and effective
magnetic moments in paramagnetic state of Mn and Ni atoms were
estimated and the influence of delocalization effects on magnetism
in these alloys was discussed. The experimental results obtained
show that the shift of martensitic transition temperature depends
weakly on composition. The values of this shift are in good
correspondence with Clapeyron-Clausius formalism taking into
account the experimental data on latent heat at
martensite-austenite transformations.
\end{abstract}

\pacs{64.70.Kb, 75.30.Cr, 75.50.Cc}


\maketitle

\section{Introduction}

In Ni$_2$MnGa Heusler alloy, a structural transformation from
cubic austenitic to tetragonal martensitic phase is observed upon
cooling. The interest in the study of Ni$_2$MnGa-based alloys has
mainly been conditioned by the fact that the martensitic phase in
these alloys is ferromagnetic. The combination of ferromagnetic
ordering and martensitic transformation allows realization of
magnetically driven shape memory effect, which expands
considerably the area of technical applications of this effect.

Despite a large number of experimental and theoretical studies,
many fundamental aspects of Ni$_2$MnGa-based alloys are not
clearly understood yet. For instance, magnetic properties of
thoroughly studied Ni$_{2+x}$Mn$_{1-x}$Ga system were not
sufficiently clarified. For these alloys, the compositional
dependencies of Curie temperature $T_C$ and martensite-austenite
transformation temperature $T_m$ were determined but the
temperature and compositional dependencies of magnetization have
been not investigated in details. In particular, no systematic
study was performed on the jump of magnetization at martensitic
transition, which determines the shift of $T_m$ under external
magnetic field. Besides, the exchange interaction parameters have
been not estimated for these alloys. All these factors are
important to get a better insight into physical mechanisms,
underlying the magnetically driven shape memory effect. This paper
deals with a systematic study of magnetic properties of
Ni$_{2+x}$Mn$_{1-x}$Ga ($0 \le x \le 0.19$) alloys.

\section{Crystal structure and magnetic properties of Ni$_{2+x}$Mn$_{1-x}$Ga
system}

The high-temperature austenitic phase of Ni$_{2+x}$Mn$_{1-x}$Ga
Heusler alloys has a cubic structure of Fm$\bar{3}$m space group.
A structural transition to a modulated tetragonal ($c/a < 1$)
phase is observed in these alloys on cooling. It is worth noting
that the crystal structure and space group of the low-temperature
phase is still a subject of controversy (see, for example,
Refs.~\onlinecite{2-w,3-p}) which is aggravated by a compositional
dependence of the crystal structure of martensite. Thus, for
example, recent results of high-resolution neutron
diffraction~\cite{4-b} give ground to conclude that for the
stoichiometric Ni$_2$MnGa composition the martensitic phase, being
considered for a long time as a tetragonal, has indeed an
orthorhombic symmetry of Pnnm space group. The structural
martensitic transformation in Ni$_2$MnGa-based Heusler alloys was
described as driven by a band Jahn-Teller effect.~\cite{5-b,6-f}

The martensitic transformation temperature $T_m$, which is about
200~K in stoichiometric Ni$_2$MnGa, linearly increases with $x$ in
Ni$_{2+x}$Mn$_{1-x}$Ga alloys and reaches about 330~K at $x = 0.18
- 0.19$ (Ref.~\onlinecite{7-v}). The alloys with a higher Ni
content were not studied so far. Note that different values of
$T_m$ are given in literature, indicating probably the sensitivity
of physical properties of these alloys to structural
disorder~\cite{8-h,9-k} and/or deviations from the nominal
composition.

The Ni$_{2+x}$Mn$_{1-x}$Ga alloys are ferromagnetic at low
temperatures. The Curie temperature $T_C$ is about 370~K in
stoichiometric composition ($x = 0$). $T_C$ approximately linearly
decreases with increasing Ni content, so that for $x = 0.18-0.19$
Curie temperature merges with the martensitic transformation
temperature $T_m$. Hence, the alloys with $x = 0.18 - 0.19$
experience a structural (martensitic) transition from paramagnetic
austenite to ferromagnetic martensite. At the same time, the
magnetic state of the alloys with a lower Ni content does not
change during martensitic transformation and both austenitic and
martensitic phases are ferromagnetic. The martensitic
transformation, however, influences the magnetic parameters of
these alloys and reveals itself in a sharp change of magnetic
anisotropy and magnetization saturation.~\cite{10-t}

The neutron diffraction measurements of stoichiometric
composition~\cite{5-b,10a-w} show that the magnetic moment is
localized mainly on Mn atoms. The reported values of the Mn
magnetic moment range from 3.8 to 4.2 $\mu_B$. The magnetic moment
of Ni atoms is considerably smaller, about 0.2-0.4 $\mu_B$. The
concentration dependence of magnetic moment in
Ni$_{2+x}$Mn$_{1-x}$Ga alloys has been not reported. It is known
only that magnetization saturation decreases with increasing
$x$.~\cite{11-w,12-a}

\section{Samples preparation and measurements}

Polycrystalline samples of Ni$_{2+x}$Mn$_{1-x}$Ga alloys were
prepared by a conventional arc-melting method in the atmosphere of
spectroscopically pure argon gas. The samples were homogenized at
1050~K for 9 days with subsequent quenching in ice water. For the
measurements of physical properties those samples were used whose
weight loss during arc-melting was less than 0.2\%. The
measurements of magnetic properties were performed on samples with
$x$ = 0, 0.04, 0.08, 0.12, 0.16, and 0.19; some measurements were
also done on the sample with $x = 0.02$.

The magnetization up to 5~T was measured in a temperature range 5
-- 700~K by a SQUID magnetometer "Quantum Design"; it was also
measured by a vibrating sample magnetometer (VSM) in magnetic
fields up to 1.8~T. Additionally, measurements in pulsed magnetic
fields up to 10~T were performed. Spontaneous magnetization $M_s$
at low temperatures was determined by linear extrapolation of
$M(H)$ dependencies from high fields. $M_s$ in the vicinity of
Curie temperature, where $M(H)$ dependencies are non-linear, was
estimated by Belov-Arrott method for second-order magnetic phase
transitions. Using this method, the Curie temperatures were
determined for every alloys except the $x = 0.19$ sample, where
the ferromagnetic-paramagnetic transition is a first-order phase
transition. The paramagnetic susceptibility of the alloys was
defined from $M(T)$ dependencies measured above $T_C$ up to 700~K
in a magnetic field of 0.2~T.

The latent heat of martensitic transition was determined from
differential scanning calorimetry, performed by a Pyris-1 DSC
equipment.

\begin{figure}[t]
\includegraphics[width=\columnwidth]{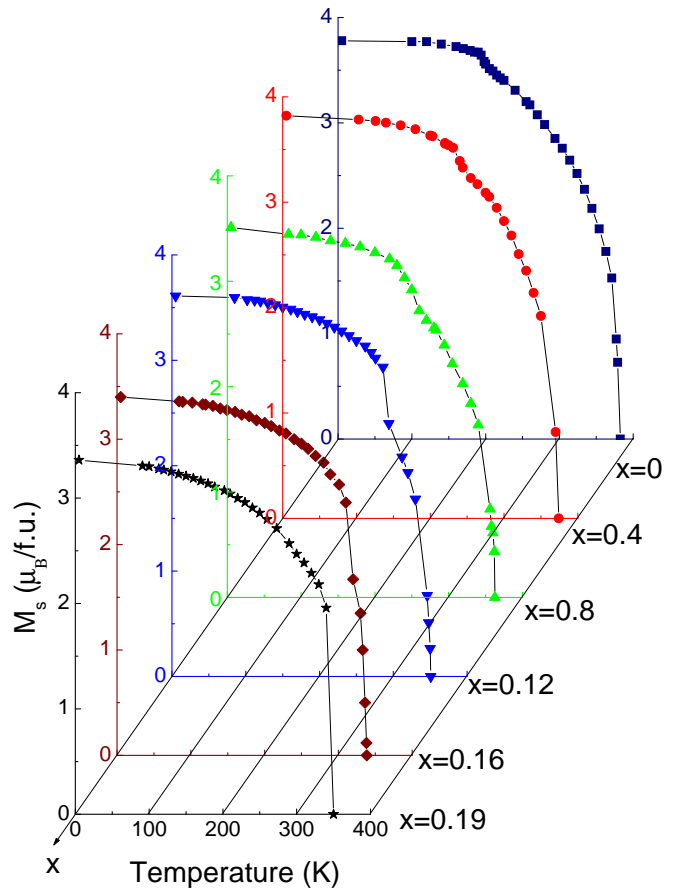}
\caption{Temperature dependencies of spontaneous magnetization of
the Ni$_{2+x}$Mn$_{1-x}$Ga alloys.}
\end{figure}

\section{Experimental results}

Temperature dependencies of spontaneous magnetization $M_s$ of the
Ni$_{2+x}$Mn$_{1-x}$Ga alloys are shown in Fig.~1. It is seen that
$M_s$ gradually decreases with increasing temperature and exhibits
a pronounced change (smeared jump) when approaching a certain
temperature $T_m$. This jump in magnetization is caused, as has
been shown in numerous studies,~\cite{10a-w,13-k,14-g} by a
structural phase transition from martensite to austenite. As
evident from these measurements, the austenitic phase is
ferromagnetic above $T_m$ for $x < 0.19$, while in the $x = 0.19$
alloy the transformation from martensite to austenite is
accompanied by a transition from ferromagnetic to paramagnetic
state.


\begin{figure*}
\begin{minipage}[t]{\columnwidth}
\setcaptionwidth{8cm}
 \centering
\includegraphics[width=6cm]{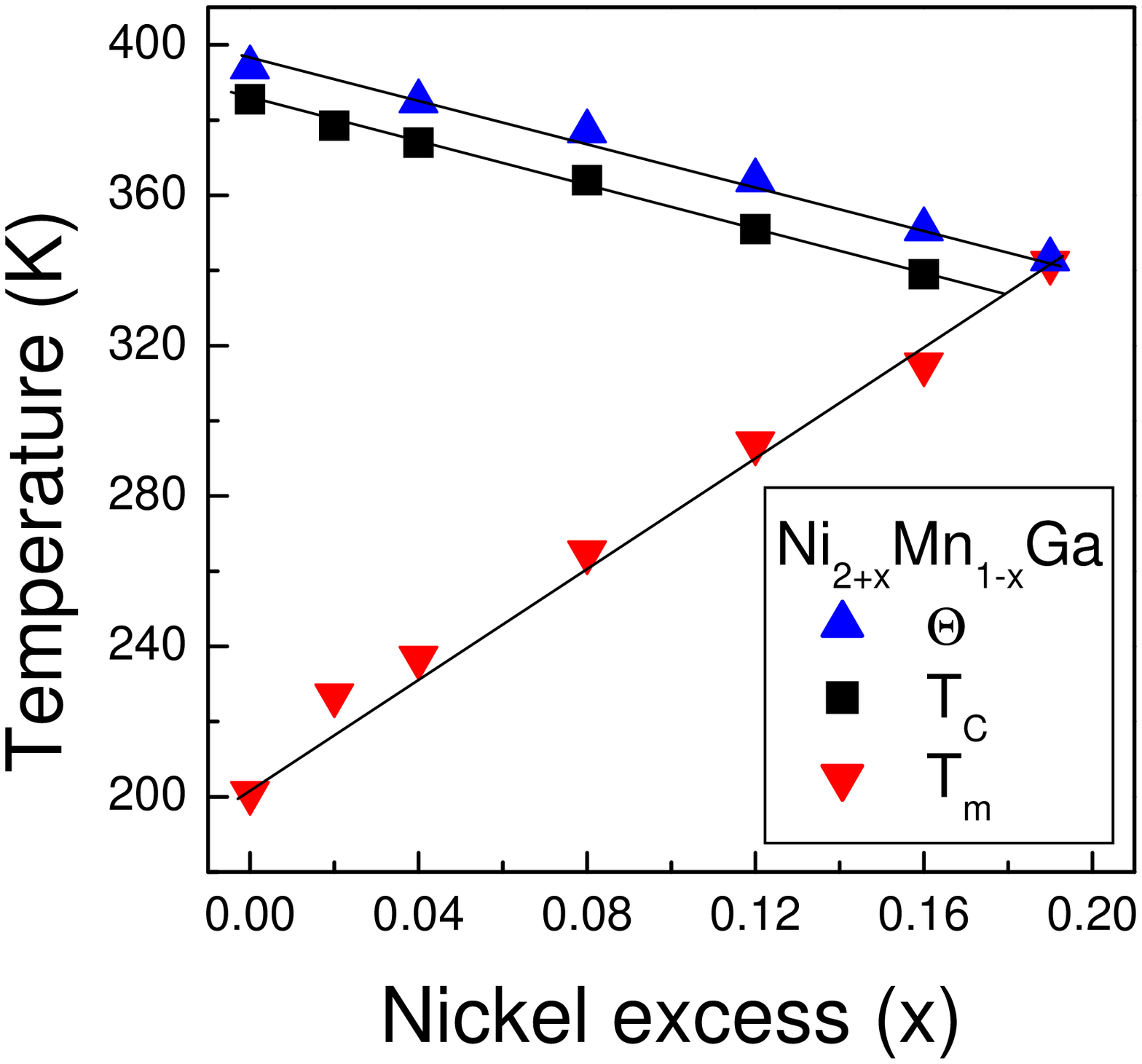}
\caption{Compositional dependencies of martensitic transformation
temperature $T_m$, Curie temperature $T_C$ and paramagnetic Curie
temperature $\Theta$.}
\end{minipage}
\begin{minipage}[t]{\columnwidth}
\centering
\includegraphics[width=6cm]{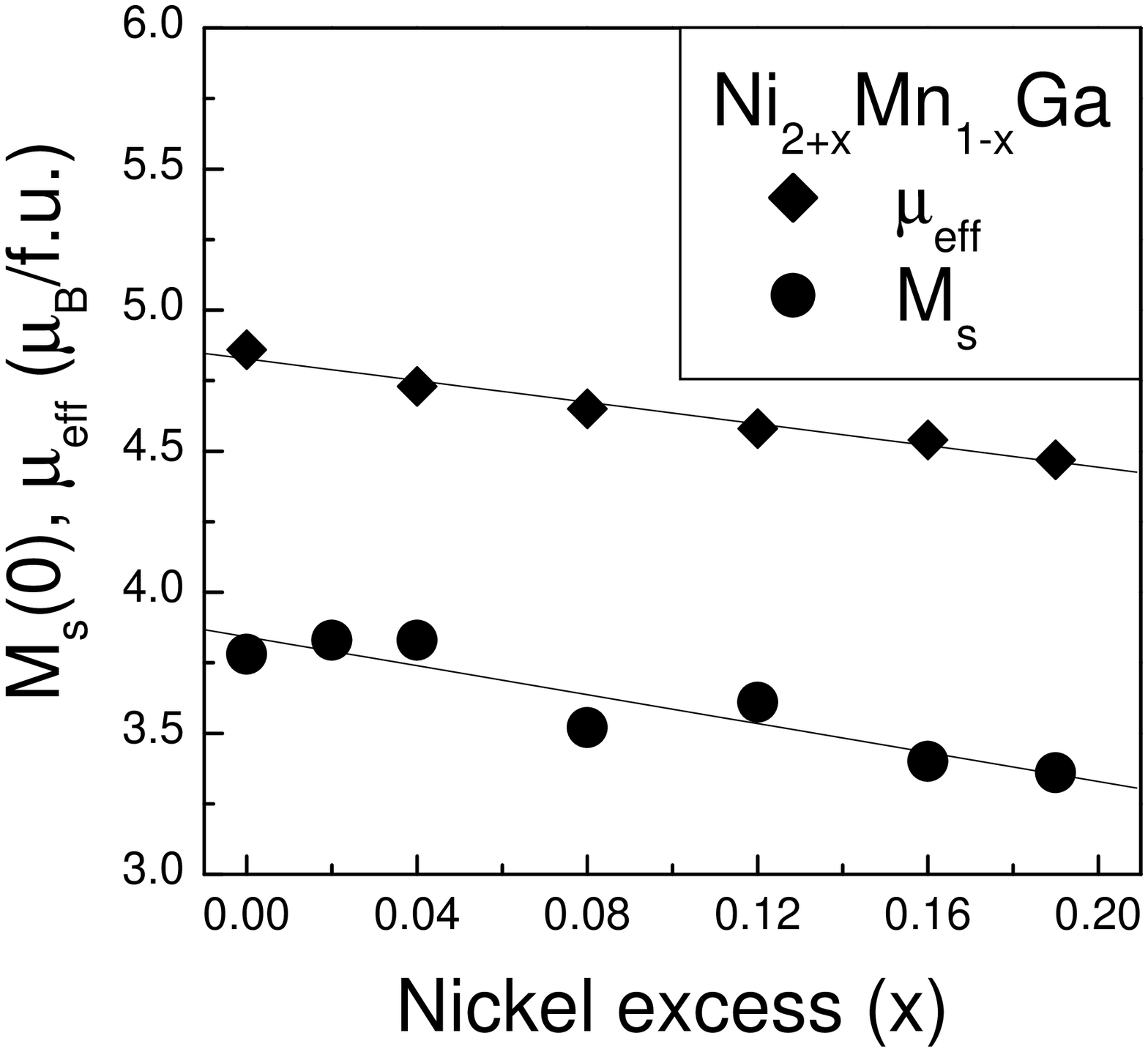}
\caption{Compositional dependencies of saturation magnetic moment
$M_s(0)$ and effective magnetic moment $\mu_{\mathrm{eff}}$ of the
Ni$_{2+x}$Mn$_{1-x}$Ga alloys.}
\end{minipage}

 \centering
 \vspace{0.5cm}
\includegraphics[width=12cm]{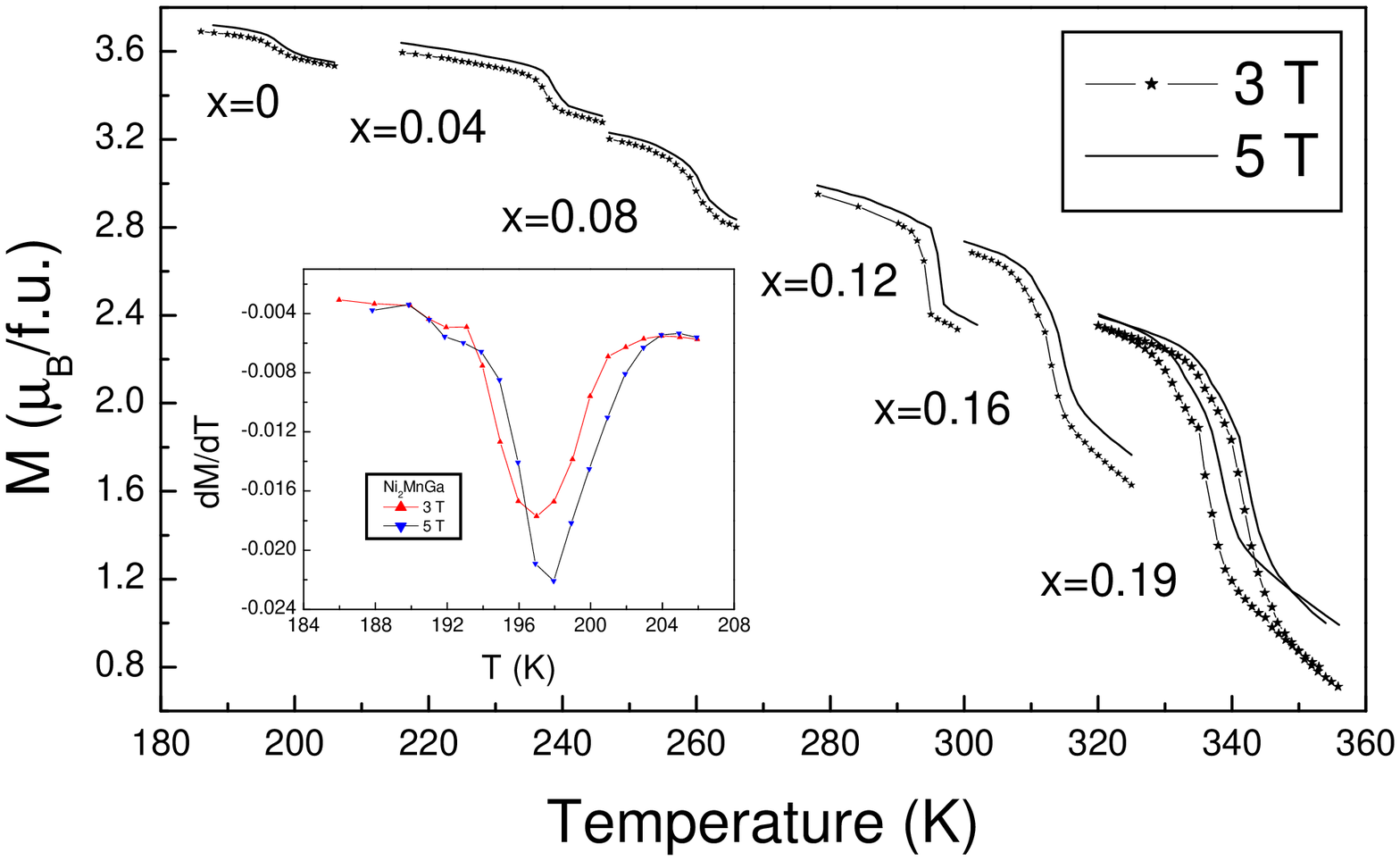}
\caption{Magnetization jump at the martensitic transition in
magnetic fields of 3~T (dashed line) and 5~T (solid line). For
compositions $0 \le x \le 0.16$ $M(T)$ dependencies were measured
upon heating. For the composition $x = 0.19$ a temperature
hysteresis loop of the magnetization observed at martensitic
transition is shown. The inset shows temperature derivatives of
magnetization for Ni$_2$MnGa measured in magnetic fields 3 and
5~T.}
\end{figure*}

The compositional dependence of the martensitic transformation
temperature $T_m$ is shown in Fig.~2. This figure also shows the
dependence of ferromagnetic ordering temperature $T_C$ on Ni
content $x$. It is seen that both these dependencies are
practically linear with $T_m$ increasing and $T_C$ decreasing with
Ni content. These temperatures merge in a range of $x = 0.18 -
0.19$. The phase diagram of the Ni$_{2+x}$Mn$_{1-x}$Ga system
obtained in the present study is in good agreement with previously
found one.~\cite{7-v,15-b}

The magnetic moment of these alloys, $M_s(0)$, was obtained by
extrapolation of $M_s(T)$ to 0~K. It was found that $M_s(0)$
approximately linearly decreases at substitution of Mn by Ni, as
is shown in Fig.~3. The value of magnetic moment in the
stoichiometric Ni$_2$MnGa appears to be close to those reported in
others studies.~\cite{5-b,10a-w,11-w}

The $M_s(T)$ dependencies in Ni$_{2+x}$Mn$_{1-x}$Ga alloys, shown
in Fig.~1, evidence that the change of spontaneous magnetization
at martensite-austenite transformation increases with Ni content.
A jump of magnetization at this transition is also observed in
magnetic fields larger than the saturation field, as is shown in
Fig.~4. The compositional dependencies of the magnitudes of
magnetization jump $\Delta M$ measured in various magnetic fields
are shown in Fig.~5.

As can be seen from Fig.~4, with increasing magnetic field the
magnetization jumps shift to higher temperatures. This is due to
the influence of magnetic field on martensitic transformation
temperature. It follows from these measurements that the shift
$\Delta T$ of $T_m$ under magnetic field increases weakly  with Ni
content (see Table~1).



The influence of a magnetic field on the martensite-austenite
transition temperature was studied only for $x = 0$
(Ref.~\onlinecite{14-g}) and $x = 0.18 - 0.19$
(Refs.~\onlinecite{15-b,16-k,17-f}) compositions. For the
stoichiometric composition the shift of $T_m$ under magnetic field
was estimated as $dT_m/dH \approx 0.2$~K/T.~\cite{14-g} For the $x
= 0.18$ and $x = 0.19$ compositions, $dT_m/dH \approx 1$~K/T was
reported in Refs.~\onlinecite{15-b,17-f}, whereas in
Ref.~\onlinecite{16-k} this quantity was estimated as 3.5~K/T. It
is worth noting that the shift of $T_m$ is determined with a
significant error. This is caused mainly by the fact that the jump
of magnetization at martensitic transformation is broad which
makes difficult correct determination of $T_m$ temperature.
Besides, martensitic transformation is a first-order structural
phase transition and is characterized by a temperature hysteresis.
Therefore, $T_m$ temperature can differ from the temperature at
which the jump of magnetization occurs. The most correct method to
determine $T_m$ is to determine this temperature as the average of
the temperatures, at which magnetization jump is observed on
cooling and heating, respectively. In the present study, $T_m$ was
determined as a temperature of the magnetization jump while
heating the sample. The temperature hysteresis loop was measured
for the $x = 0.19$ sample. It was found (see Fig.~4) that $T_m$
determined at increasing temperature differs from $T_m$ estimated
from averaging of measurements in hysteretic regime by 2--3 K. The
width of the temperature hysteresis loop is approximately the same
in different magnetic fields, so the additional error in
determination of the shift of $T_m$ caused by a magnetic field
does not exceed 0.3~K.

\begin{table}[t]
\caption{Theoretical and experimental values of the shift $\Delta
T$ of the martensitic transition temperature $T_m$ in a magnetic
field $\Delta H$ = 2~T for Ni$_{2+x}$Mn$_{1-x}$Ga.}
\begin{ruledtabular}
\begin{tabular}{cccccc}

     &         &      &      &  $\Delta T$ (K) & $\Delta T$ (K) \\

$x$  &  $Q$ (J/mol)  &  $T_m$ (K)  & $\Delta M$ ($\mu_B$) &
($\Delta H$ = 2 T) & ($\Delta H$ = 2 T) \\
     &         &      &      & theory  & experiment \\

\hline

0 &  270 & 201 & 0.1 & $0.82\pm0.2$  &  $0.8\pm0.5$  \\

0.04  &  600 & 237 & 0.17 &   $0.75\pm0.2$  &  $0.95\pm0.5$ \\

0.08  &  910 & 265 & 0.28 &  $0.92\pm0.2$   &  $0.95\pm0.5$  \\

0.12  &  1250 &  294 & 0.41 &   $1.07\pm0.2$  &  $1.10\pm0.5$ \\

0.16  &  1710 &   315 & 0.62 &   $1.28\pm0.2$  &   $1.30\pm0.5$ \\

0.19  &  2260 &   342 & 0.96 &   $1.62\pm0.2$  &  $1.60\pm0.5$  \\


\end{tabular}
\end{ruledtabular}
\end{table}

\begin{figure}[t]
\includegraphics[width=7cm]{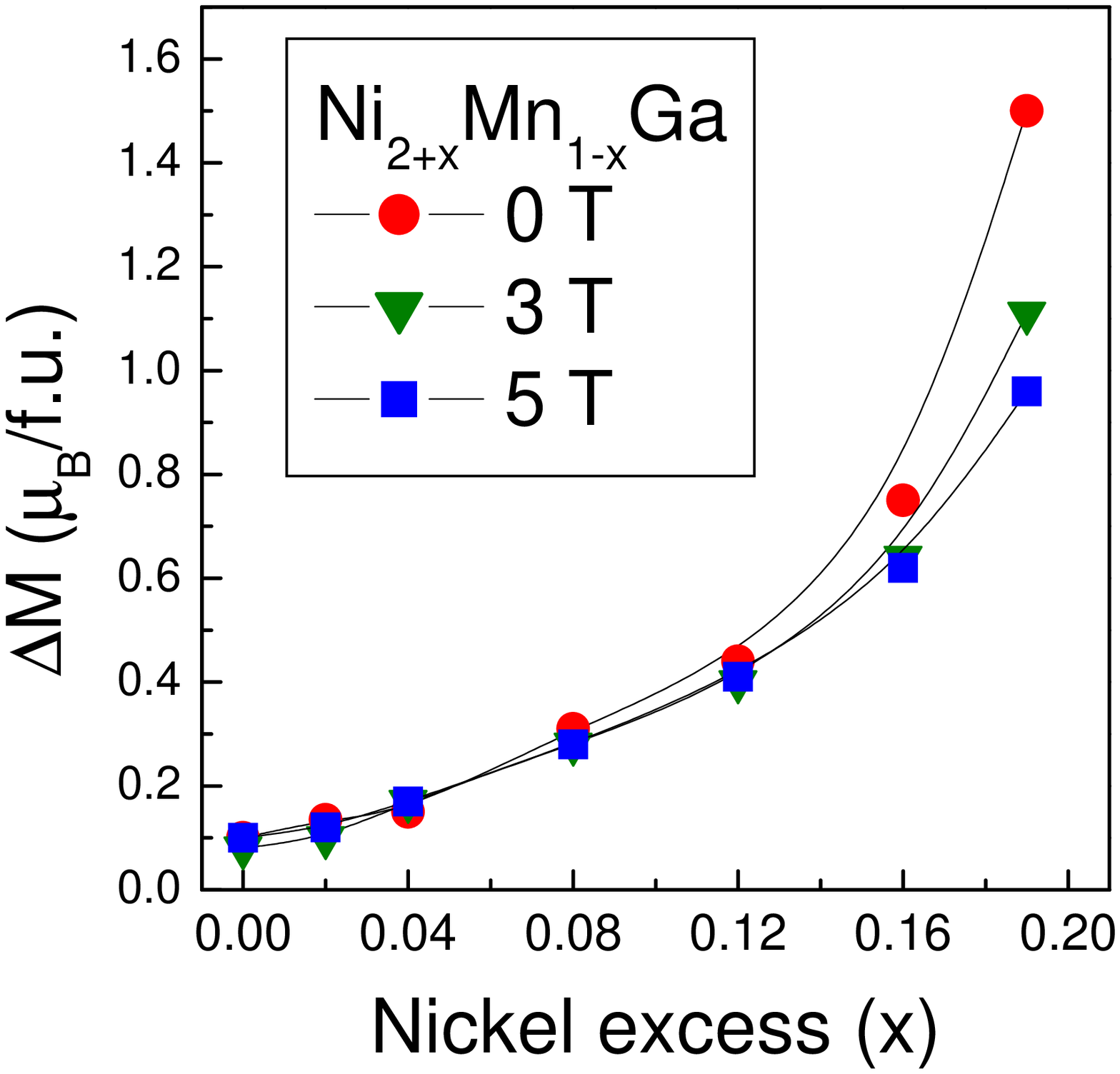}
\caption{The magnetization jump at martensitic transition in
various magnetic fields as a function of Ni concentration in the
Ni$_{2+x}$Mn$_{1-x}$Ga alloys.}
\vspace{0.5cm}
\includegraphics[width=7cm]{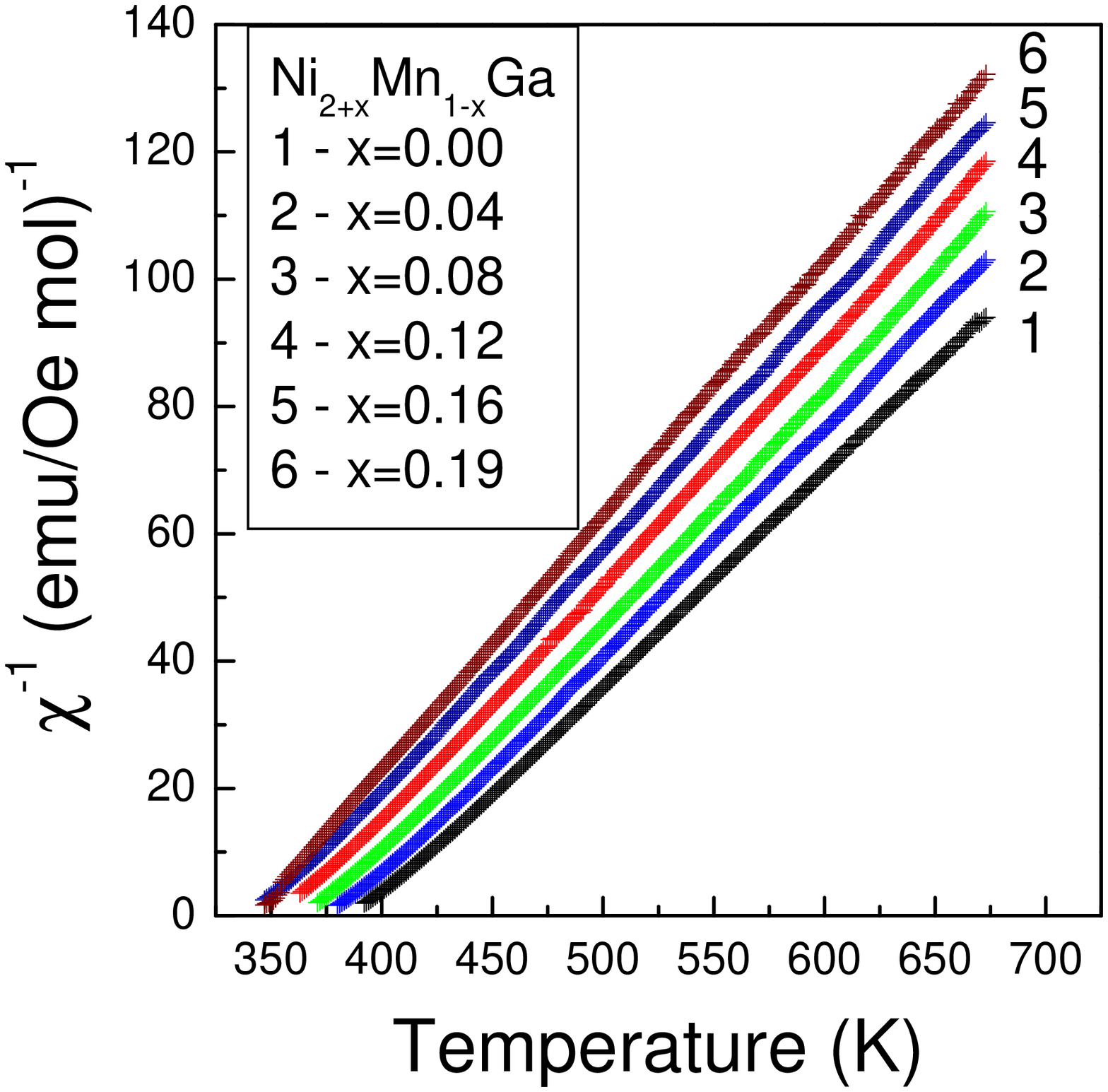}
\caption{Temperature dependencies of paramagnetic susceptibility
in Ni$_{2+x}$Mn$_{1-x}$Ga alloys.}
\end{figure}


Temperature dependencies of reciprocal paramagnetic susceptibility
of the Ni$_{2+x}$Mn$_{1-x}$Ga system are shown in Fig.~6. In the
temperature range studied, the susceptibility follows Curie-Weiss
law. The compositional dependencies of paramagnetic Curie
temperature $\Theta$ and effective magnetic moment
$\mu_{\mathrm{eff}}$ are shown in Figs.~2 and 3, respectively.
Clearly, both these parameters decrease monotonously with
increasing $x$. The paramagnetic susceptibility was measured
earlier in stoichiometric Ni$_2$MnGa alloy only.~\cite{10a-w,18-k}
The values of $\Theta$ and $\mu_{\mathrm{eff}}$ obtained are
somewhat larger than the reported previously. This difference can
be due to the fact that the present measurements were performed in
a wider temperature interval.

The compositional dependence of the latent heat $Q$ of the
martensite-austenite phase transition is shown in Fig.~7.
Evidently, $Q$ strongly increases with increasing $x$. These
results are in good agreement with recently published
ones.~\cite{19-k}

\section{Discussion}

Based on the compositional dependencies of saturation magnetic
moment and effective magnetic moment of Ni$_{2+x}$Mn$_{1-x}$Ga
alloys (Fig.~3), the magnetic moments and effective magnetic
moments of Mn and Ni atoms were calculated from the equations

\begin{equation}
M_s(0)=(1-x)\mu_{\mathrm{Mn}} + (2+x)\mu_{\mathrm{Ni}}
\end{equation}

\begin{equation}
\mu_{\mathrm{eff}} = \sqrt{(1-x)\mu_{\mathrm{eff}\,\mathrm{Mn}}^2
+ (2+x)\mu_{\mathrm{eff}\,\mathrm{Ni}}^2}\;.
\end{equation}

\noindent The results of these calculations are presented in
Table~2. The obtained values of the magnetic moments of the
constituting atoms are in good accordance with the results of
neutron diffraction and nuclear magnetic resonance studies for the
stoichiometric composition. It was shown~\cite{4-b,5-b,20-o} that
in Ni$_2$MnGa the Mn magnetic moment is about $2.84-3.41 \mu_B$
and the Ni magnetic moment is about $0.3-0.41 \mu_B$.

\begin{table}[t]
\caption{Magnetic moments $\mu$ and effective magnetic moments
$\mu_{\mathrm{eff}}$ of Mn and Ni atoms.}
\begin{ruledtabular}
\begin{tabular}{ccccc}

  & $\mu$ $(\mu_{B})$ & $\mu_{\mathrm{eff}}$ $(\mu_{B})$ &
$\mu_{\mathrm{eff}}^{\mathrm{loc}}$ $(\mu_{B})$ &
$\mu_{\mathrm{eff}}^{\mathrm{loc}}/\mu_{\mathrm{eff}}$   \\

\hline

Mn & $2.99\pm0.32$ & $4.43\pm0.13$ & $3.86\pm0.14$ & $0.87\pm0.11$
\\

Ni & $0.43\pm0.14$ & $1.35\pm0.18$ & $1.05\pm0.21$ & $0.77\pm0.10$
\\
\end{tabular}
\end{ruledtabular}
\end{table}

Note that these calculations are based on assumptions that the
magnetic moments of constituting atoms does not change with
deviations from stoichiometry and that the Ni atoms possess
similar moments in different crystallographic sites. In general
this is not the case, because magnetism of Heusler alloys is
described in a band model. It means that the values of magnetic
moments depend on density of states at Fermi level and on the
exchange splitting parameter, being therefore the concentration
and structure dependent values. As has been noted in
Ref.~\onlinecite{21-m}, in Ni$_2$MnX Heusler alloys the distance
between the atoms is sufficiently large so that direct overlap of
electron orbitals is negligible and the delocalization effects are
of secondary importance. Therefore, in the first approximation a
localized moments model is applicable for the description of
magnetic properties of these alloys. However, from the results of
magnetic and nuclear magnetic resonance measurements of
Ni$_2$MnGa~\cite{20-o} it was concluded that in this alloy the Mn
magnetic moments are mainly localized, while Ni magnetic moments
are essentially delocalized.

The character of magnetism can be judged from the comparison of
the magnetic moments of the constituting atoms and their effective
magnetic moments (see Table~2).

In the model of localized magnetic moments for the spin-only state
(orbital moment is quenched) the interrelation between effective
magnetic moment and the moment in the magnetically ordered state
is given by Wohlfarth-Rhodes equation

\begin{equation}
\mu_{\mathrm{eff}}^{\mathrm{loc}} = \sqrt{\mu(\mu+2)}
\end{equation}

\noindent In the band model the value of
$\mu_{\mathrm{eff}}^{\mathrm{loc}}$ calculated from Eqn.~(3)
should be smaller than the experimental value of the effective
moment due to the influence of delocalization effects. The values
of $\mu_{\mathrm{eff}}^{\mathrm{loc}}$ for the Mn and Ni
subsystems are given in Table~2. As evident from these data, for
both Mn and Ni subsystems $\mu_{\mathrm{eff}}$ and
$\mu_{\mathrm{eff}}^{\mathrm{loc}}$ are close to each other,
although in both cases $\mu_{\mathrm{eff}}^{\mathrm{loc}}$ is
slightly smaller than $\mu_{\mathrm{eff}}$. Within the
experimental error of the measurements the
$\mu_{\mathrm{eff}}^{\mathrm{loc}}/\mu_{\mathrm{eff}}$ ratio is
the same for both Mn and Ni subsystems. Thus, present experimental
data do not suggest that the Ni subsystem is more delocalized that
the Mn one.

\begin{figure}[t]
\includegraphics[width=7cm]{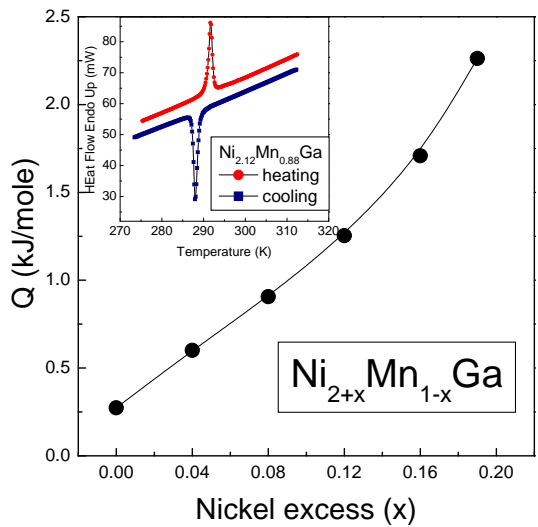}
\caption{Compositional dependence of the latent heat of the
martensitic transition. The inset shows an example of the
differential scanning calorimetry measurements.}
\end{figure}

It should be understood, however, that the magnetic moments in the
magnetically ordered state were determined in the martensitic
phase, whereas the effective magnetic moments were calculated from
the paramagnetic susceptibility measured in the austenitic state.
It makes no difference if magnetism is described in the localized
model, because in this case the magnitude of magnetic moment
depends weakly on the crystallographic environment. In the band
model, magnetic moments depend on the degree of overlap of
electron orbitals, which changes at structural transformation.
Because of this, a possibility that the magnetic moments will
change at structural transition must not be ruled out. The
qualitative arguments given above are supported by the
experimental data reported in Ref.~\onlinecite{20-o}, which
indicate that the magnetic moment of Mn is the same in austenitic
and martensitic phases, whereas magnetic moment of Ni in
austenitic phase is larger than that in the martensitic phase. The
latter observation is conditioned by a higher density of states of
Ni at the Fermi level in austenitic state than that in the
martensitic state, as electronic structure calculations have
revealed.~\cite{6-f}

As evident from Fig.~2, Curie temperature of the austenitic phase
decreases at substitution of Mn for Ni. This is due to the fact
that this substitution leads to an increase in the number of atoms
with smaller magnetic moments. Similar tendency takes place
presumably for a virtual Curie temperature of the martensitic
phase. This follows from the observation that in the
low-temperature martensitic phase magnetization of the
Ni$_{2+x}$Mn$_{1-x}$Ga alloys with a higher $x$ decreases more
rapidly with increasing temperature.

The magnetization data shown in Fig.~1 allow estimation of virtual
Curie temperature of the martensitic phase. Fig.~8 shows
temperature dependencies of reduced spontaneous magnetization $m =
M_s(T)/M_s(0)$ of the alloys as a function of reduced temperature
$t = T/T_C$. It is seen that the magnetization of austenitic phase
and the magnetization of martensitic phase change with temperature
in different way, whereas the reduced magnetizations of these
phases are similar for different compositions. It can be assumed
that the difference in $m(t)$ of martensite and austenite is due
to the difference in their Curie temperatures. Comparing $m(t)$
dependencies of martensitic and austenitic phases, it is possible
to reconstruct the virtual Curie temperature of martensitic phase,
which is shown by the solid line in Fig.~8. It appears to be 17\%
higher than the Curie temperature of austenitic state. This value
is twice as large as that obtained from phenomenological Landau
theory.~\cite{22-c}

\begin{figure}[t]
\includegraphics[width=7cm]{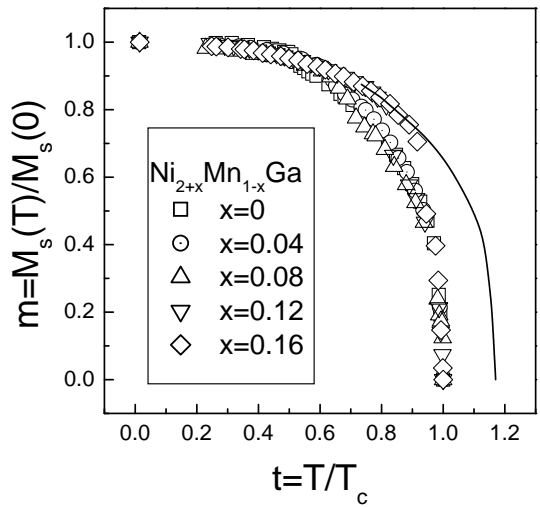}
\caption{Reduced magnetization $m=M_s(T)/M_s(0)$ as a function of
reduced temperature $t=T/T_C$ for the Ni$_{2+x}$Mn$_{1-x}$Ga
alloys. The solid line is reduced magnetization of virtual
martensitic phase.}
\end{figure}

The larger value of Curie temperature of martensite as compared to
Curie temperature of corresponding austenite is due to changes in
interatomic distances and in overlap of electronic orbitals. As
evident from the analysis of experimental data, this effect cannot
be attributed solely to a change in the unit cell volume at
martensitic transformation. Indeed, a study of the influence of
hydrostatic pressure on Curie temperature $T_C$ and martensitic
transformation temperature $T_m$ of stoichiometric
Ni$_2$MnGa~\cite{18-k} has shown that the exchange interaction of
the austenite increases with decreasing unit cell volume. At the
same time, it is known~\cite{23-m} that the unit cell volume of
martensite is larger than that of austenite. Therefore, it seems
likely that the primary role in martensitic transformation in
Ni$_2$MnGa Heusler alloys belongs to the crystal lattice
distortions. Such a mechanism of the influence of a structural
transition on exchange interaction in intermetallic compounds
Gd$_5$(Si$_x$Ge$_{1-x}$)$_4$ was discussed recently in
Ref.~\onlinecite{24-c}.

As evident from Fig.~5, the magnitude of magnetization jump
$\Delta M$ at structural transition strongly increases with Ni
content. This is caused by the fact that the increase of $x$ leads
to the increase of $T_m$. Under these circumstances, the
difference between magnetizations of martensite and austenite at
$T_m$ increases as well. It is also seen from Fig.~5 that the
magnetization jumps $\Delta M$ at $T_m$ diminishes at increasing
magnetic field, which is the most pronounced at high $x$. The
behavior of $\Delta M$ in the alloys with a small $x$ results from
the fact that the martensitic transformation in these alloys
occurs at temperatures far below Curie temperature $T_C$ and
therefore the influence of a magnetic field on magnetization is
weak. In the alloys with a large $x$, $T_m$ is close to $T_C$ of
the austenitic phase and the external field strongly affects
magnetization of this phase, whereas magnetization of the
martensitic phase depends weakly on the magnetic field.

It has been already mentioned that the temperature of structural
transition shifts to higher temperatures upon application of a
magnetic field. Such behavior is governed by the influence of
Zeeman energy, which stabilizes martensitic phase with a larger
magnetization. Experimental data on the shift of $T_m$ are
presented in Table~1. These results indicate that for the alloys
studied the shift is rather small (1--2~K as the magnetic field
changes for 2~T) and slightly enhances with increasing Ni content.
The table also contains theoretical estimation of the shift of
$T_m$ in magnetic field, derived from a thermodynamical
Clapeyron-Clausius relation for first-order phase transitions:

\[
\Delta T = \Delta MHT_m/Q\;.
\]

\noindent The agreement between experimental and theoretical
values can be considered as satisfactory.

\section*{Acknowledgements}

This study was supported by RFBR grants No. 02-02-16636 and
03-02-17443. Work at Argonne was supported by US Department of
Energy, BES Materials Sciences under contract W-31-109-ENG-38. One
of the authors (VVK) gratefully acknowledges the Japan Society for
the Promotion of Science for a Postdoctoral Fellowship Award.

\end{document}